\documentclass[sigconf]{acmart}

\def\BibTeX{{\rm B\kern-.05em{\sc i\kern-.025em b}\kern-.08emT\kern-.1667em\lower.7ex\hbox{E}\kern-.125emX}}

\usepackage{booktabs} 
\usepackage{multirow}
\usepackage{siunitx}
\usepackage{subcaption}
\usepackage[most]{tcolorbox}
\usepackage[export]{adjustbox}
\usepackage{hyperref}
\usepackage{algorithm, algpseudocode}
\usepackage{hyperref}
\usepackage{placeins}
\usepackage{graphicx}
\usepackage{cleveref}

\acmConference[ASE 2022]{37th IEEE/ACM International Conference on Automated Software Engineering}{xx - xx Month, 2022}{United States}
\acmYear{2022}
\copyrightyear{2022}

\acmPrice{xx.00}

\begin{document}

\title{A Drift Handling Approach for Self-Adaptive ML Software in Scalable Industrial Processes}

\author{Firas Bayram}
\orcid{0000-xx-xx-xx}
\affiliation{%
  \institution{Department of Mathematics and Computer Science, Karlstad University, Universitetsgatan 2, Karlstad, 651 88, Sweden}
}
\email{Firas.Bayram@kau.se}

\author{Bestoun S. Ahmed}
\orcid{1234-5678-9012}
\affiliation{%
  \institution{Department of Mathematics and Computer Science, Karlstad University, Universitetsgatan 2, Karlstad, 651 88, Sweden}
}
\email{bestoun@kau.se}

\author{Erik Hallin}
\affiliation{%
  \institution{Uddeholms AB, Uvedsv{\"a}gen, Hagfors, 683 33, V{\"a}rmlands l{\"a}n, Sweden}
}
\email{erik.hallin@uddeholm.com}

\author{Anton Engman}
\affiliation{%
  \institution{Uddeholms AB, Uvedsv{\"a}gen, Hagfors, 683 33, V{\"a}rmlands l{\"a}n, Sweden}
}
\email{anton.engman@uddeholm.com}

\renewcommand{\shortauthors}{Bayram F, et al.}

\begin{abstract}
Most industrial processes in real-world manufacturing applications are characterized by the scalability property, which requires an automated strategy to self-adapt machine learning (ML) software systems to the new conditions. In this paper, we investigate an Electroslag Remelting (ESR) use case process from the Uddeholms AB steel company. The use case involves predicting the minimum pressure value for a vacuum pumping event. Taking into account the long time required to collect new records and efficiently integrate the new machines with the built ML software system. Additionally, to accommodate the changes and satisfy the non-functional requirement of the software system, namely adaptability, we propose an automated and adaptive approach based on a drift handling technique called importance weighting. The aim is to address the problem of adding a new furnace to production and enable the adaptability attribute of the ML software. The overall results demonstrate the improvements in ML software performance achieved by implementing the proposed approach over the classical non-adaptive approach.
\end{abstract}

%
%

%
\keywords{ML software adaptability, automated industrial process, Electroslag Remelting (ESR),  non-functional requirements, changing environments, concept drift}

\maketitle


\section{Introduction}
In production systems, machine learning (ML) is an integral part of a more extensive software system that performs the overall desired functionalities \cite{lwakatare2019taxonomy}. With its powerful forecasting tools, the ML system has been widely used in real-world industrial processes, providing valuable insights for running businesses \cite{canhoto2020artificial}. Modern industries are usually characterized by different levels of scalability. Scalable contexts include the upscaling of business machinery and equipment, the acquisition of new customers, the change in market demand, or the expansion of new products \cite{accorsi2021scalability}. Therefore, industries are imposed with a significant challenge that requires performing a reasonable adaptation to work efficiently and cope with the new conditions in non-stationary environments.  

Adaptability is one of the essential non-functional requirements (NFRs), also known as extra-functional requirements \cite{panunzio2014architectural}, of any software deployed in a changing system. A general definition of NFR software adaptability is the ability of the software to deal with new conditions in response to changes in the environment \cite{subramanian2001metrics}. From an industrial perspective, adaptability is an NFR quality factor that represents the extent to which the software can exhibit business-level expandability. ML software is heavily relies on the conditions under which the system was constructed. Therefore, industrial ML software systems must inherit their characteristics from the NFR in the design phase.

Traditional ML systems accumulate their knowledge based on historical observations and trends. Thus, such systems would require their learned knowledge to be adjusted to the new conditions to maintain high performance. Moreover, due to the increased complexity of industrial systems, it would be preferable to perform such adaptations in an automated manner and minimize human intervention.  The adaptation would reduce the time required to install a new software system using a similar system already built. At the same time, automation facilitates improving the overall performance of the ML software and reducing costs by optimizing the efforts needed to integrate the new conditions \cite{lee2020bringing}. 

Modern industries in this competitive world always aim to evolve their business. A common extension for industries is the need to add new machinery for some industrial processes. Typically, this new hardware is either an upgrade to existing similar machines that perform the same functionality or introduces novel services for the industry. From an ML perspective, both situations involve the challenge of addressing the new conditions and getting the predictive system up and running as quickly as possible.

Concept drift is the machine learning situation where training and test conditions differ, which can lead to poor predictive performance in production, that is, when $P_{tr}(X,Y) \neq P_{te}(X,Y)$ \cite{bayram2022concept}. One type of concept drift is covariate shift, also known as domain adaptation, is the setting when we want to transform a pre-learned knowledge to a new condition in related domains \cite{moreno2012unifying}, that is, when $P_{tr}(X) \neq P_{te}(X)$ and $P_{tr}(Y|X) = P_{te}(Y|X)$. The discrepancy in the data distribution reflects the new conditions, whereas the identicality in the conditional distributions reflects that we are learning the same task. 


The use case of this paper is carried out at Uddeholms AB steel company in Sweden\footnote{https://www.uddeholm.com/}. The use case investigates the Electroslag Remelting (ESR) process in the steel manufacturing industry. The ESR is a continuous process that is used to remelt, refine, and solidify steel to produce ingots. The ESR process includes a sub-step, which is the vacuum pumping that extracts the oxygen from the furnace. The pumping process, which takes about 20 minutes, is stopped when the vacuum chamber reaches a certain pressure threshold. ML software predicts whether the desired threshold will be met in each pumping event. However, on average, the pumping event takes place once a day, which limits the amount of data that can be collected and used for prediction. Early identification of pumping events that will not meet expected quality is of great value to the industry; it would help save the costs of inappropriate pumping events by calling the maintenance team beforehand.

To address the aforementioned issue, we implement a covariate shift solution that enables the use of data collected from old furnaces and adapts it to be incorporated into the corresponding learning procedure of a new furnace. In other words, using drift adaptation techniques, we transfer the knowledge learned from the old furnace, which represents the source domain in our use case, to the new furnace, which represents the target domain. In this way, the scaling of the industrial process will be handled very quickly and efficiently rather than waiting for data collection from the new furnace. In this paper, we show how drift adaptation techniques improve the performance of predictive ML software when installed under new conditions.


\section{Industrial Case Overview} \label{sec:overview}

\begin{figure}
    \centering
    \includegraphics[scale=0.6]{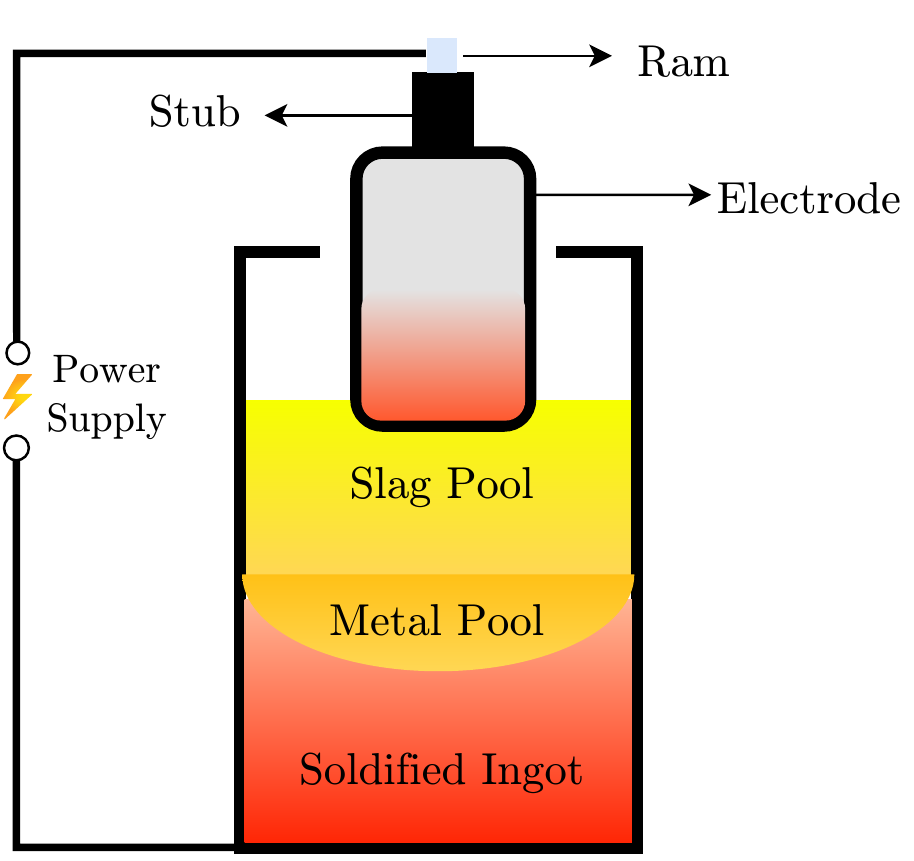}
    \caption{Electroslag Remelting Process Sketch.}
    \label{fig:sketch}
\end{figure}

The use case investigated in this paper is the industrial process of Electroslag Remelting (ESR) vacuum pumping. The vacuum process aims to extract air oxygen from the vacuum chamber. This is done for quality assurance purposes to ensure that the steel delivered is free of contaminants and gas and in an ideal condition for subsequent procedures. A schematic sketch of the ESR process is presented in Figure \ref{fig:sketch}.

The Uddeholm steel company ESR process takes place approximately once a day. And each pumping event lasts 20 minutes. Therefore, it will take almost a year to collect data for around 300 events. Pressure observations are recorded from sensors connected to the vacuum chambers. Sensors read pressure values and send them to a data storage system once every second. The storage system stores both the pressure values of the current and historical pumping events.

At the end of each pumping event, the minimum pressure value in the vacuum chamber is evaluated. Suppose that the minimum value does not reach the desired threshold. In that case, the furnace will be checked by experts for maintenance purposes, and the pumping event is re-initiated for an additional twenty minutes. Since the pressure value is proportional to the amount of gas leakage and the surface area of the vacuum chamber, the pressure value is used as an indicator to inspect the quality of the furnaces.

ML software is used to forecast the minimum pressure value based on the first observations of the pressure records that are the input of the ML model. The learning task is implemented in an automated setup for early detection of pumping issues and saves the cost of improper pumping events. Given the limited amount of data that can be collected and the use-case specifications, we developed a self-adaptation solution using drift handling approaches. The solution can be used in scalable processes and satisfy the NFR adaptability, such as adding a new furnace where there are very few (and possibly none) historical data records exist.

\section{Workflow}
\label{overview}
\begin{figure*}
    \centering
    \includegraphics[scale=0.82]{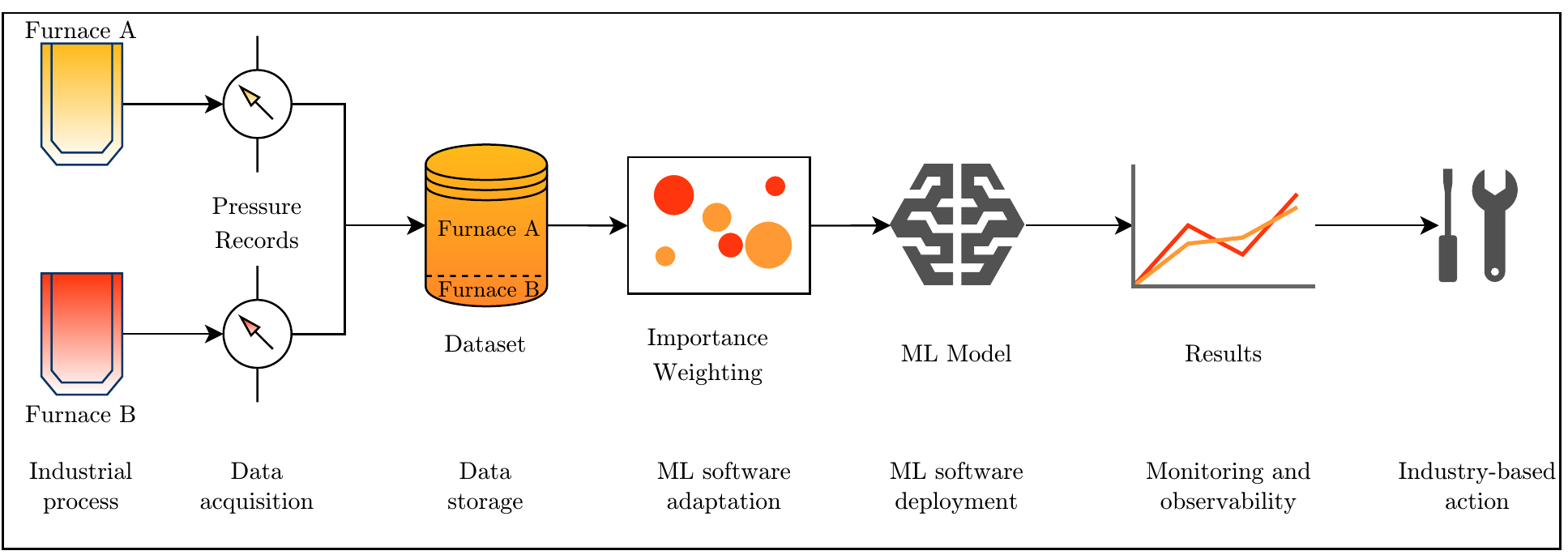}
    \caption{Automated drift-adaptive workflow for the ESR use case.}
    \label{fig:workflow}
\end{figure*}

In this section, we present the overall workflow implemented to address our ESR use case. This self-adaptive workflow aims to automatically address scalable processes in industries. The overall workflow is illustrated in Figure \ref{fig:workflow}. The evaluation is carried out on the scenario of adding new furnace(s) to the company for the vacuum pumping process. The issue arises from the lack of historical data for the new furnace, namely furnace B (the target domain), and the relatively long time required to collect the data to train the ML model. Therefore, the datasets collected from the old furnace(s), namely furnace A (the source domain), are adjusted to the new conditions and used to train the ML model.


Initially, the data were collected from furnace A and stored in the system. Once a new furnace is introduced to the ESR process, an automated process is started to re-weight the data of furnace B according to the data distribution of the pressure values registered from furnace A. The weight information is used to adapt the target data to the source data and will be used as input to the ML model for the new furnace. As illustrated in Figure \ref{fig:workflow}, the weight values are incorporated into the ML models that perform the predictive task. For the latter, we have implemented decision tree algorithms to predict the minimum pressure value of each vacuum pumping event. Despite the fact that importance weighting is a general concept and can be exploited in various ML algorithms by adjusting the corresponding loss function \cite{sugiyama2012machine}, we implemented ensemble algorithms based on decision trees which are Random Forests (RF)  \cite{breiman2001random} and XGBoost \cite{chen2016xgboost} for our solution due to their superiority in providing good results in small run time with low computational cost \cite{lin2017ensemble}. Furthermore, adding additional constraints to more complex algorithms would result in a longer average prediction time, which may violate the deadline required for the task. For the model evaluation, we used the mean absolute percentage error (MAPE) measure to monitor the performance of the ML model. MAPE is one of the most commonly used performance metrics by industrial practitioners for regression tasks due to its interpretability and independence of the scale of the variables \cite{wang2021advantages}. 

This weighting mechanism is called importance weighting, which solves the problem of covariate shift by ensuring that the data distributions of the source and target domains are aligned \cite{sugiyama2007direct}. The weights are going to be incorporated by introducing a function $w(\boldsymbol{x})$ that represents the density ratio between the target and source domain labelled data: $w(\boldsymbol{x})=\frac{p_{\mathrm{te}}(\boldsymbol{x})}{p_{\mathrm{tr}}(\boldsymbol{x})}$, where $p_{\mathrm{te}}$ is the data distribution of the target domain inputs $\left\{\boldsymbol{x}_{i}^{\operatorname{te}}\right\}_{i=1}^{n_{\mathrm{te}}}$ and $p_{\mathrm{tr}}$ is the data distribution of the source domain inputs $\left\{\boldsymbol{x}_{i}^{\operatorname{tr}}\right\}_{i=1}^{n_{\mathrm{tr}}}$. The main objective of importance weighting-based covariate shift corrections is to find the optimal density ratio function $w(\boldsymbol{x})$ such that the empirical risk under the source data distribution $R_{w}^{tr}(\ell)$ matches the empirical risk under the target data distribution $R_{w}^{te}(\ell)$ \cite{stojanov2019low}.


For covariate shift adaptation, we use the kernel mean matching method (KMM) \cite{huang2006correcting} that directly finds importance weights based on the density ratio without estimating the densities for the source and target data separately. Specifically, KMM finds the importance estimates by minimizing the Maximum Mean
Discrepancy (MMD) between $p_{\mathrm{te}}(x)$ and $w(\boldsymbol{x})p_{\mathrm{tr}(x)}$ in a Reproducing Kernel Hilbert Space (RKHS) $\Phi(x)$. KMM finds the importance weights by solving the following optimization problem \cite{sugiyama2008direct}:
$$
\min _{\left\{w_{i}\right\}_{i=1}^{n_{\mathrm{tr}}}}\left[\frac{1}{2} \sum_{i, j=1}^{n_{\mathrm{tr}}} w_{i} w_{j} K_{\sigma}\left(\boldsymbol{x}_{i}^{\mathrm{tr}}, \boldsymbol{x}_{j}^{\mathrm{tr}}\right)-\sum_{i=1}^{n_{\mathrm{tr}}} w_{i} \kappa_{i}\right]
$$
subject to $w_i\in[0, B], i=1, \ldots, n_{tr}$ and $\left|\sum_{i=1}^{n_{\mathrm{tr}}} w_{i}-n_{\mathrm{tr}}\right| \leq n_{\mathrm{tr}} \epsilon \quad$
where $K_{\sigma}\left(\boldsymbol{x}_{i}^{\mathrm{tr}}, \boldsymbol{x}_{j}^{\mathrm{tr}}\right)$ is the radial basis function (RBF) kernel, or Gaussian kernel, with width $\sigma$, $\kappa_{i}:=\frac{n_{\mathrm{tr}}}{n_{\mathrm{te}}} \sum_{j=1}^{n_{\mathrm{te}}} K_{\sigma}\left(\boldsymbol{x}_{i}^{\mathrm{tr}}, \boldsymbol{x}_{j}^{\mathrm{te}}\right)$, $B$ and $\epsilon$ are tuning parameters that control the constraints on the minimization problem. The first constraint restricts the weight interval between $0$ and $B$ to limit the influence of each input point. The second constraint ensures that the estimated function $P_{te}(\boldsymbol{x})$ is close to a probability density function (PDF) \cite{miao2015ensemble}. For parameter selection, we use the same value settings as suggested by Huang \textit{et al.} in the original paper \cite{huang2006correcting}, $B=1000$, and $\epsilon=\left(\sqrt{n_{t r}}-1\right) / \sqrt{n_{t r}}$, while the kernel width can be found using a cross-validation procedure.

\section{Evaluation Results}
In this section, we present the evaluation results after applying the self-adaptation technique introduced in the previous section. Comparisons were conducted separately on the old and new furnaces using adaptive and non-adaptive approaches. The results were monitored and recorded using different time intervals with respect to the pressure value of the vacuum pumping events. The minimum threshold is forecast every 30 seconds between the first and third minutes.

\begin{figure*}
\begin{subfigure}[b]{0.49\textwidth} 
    \centering
    \includegraphics[scale=0.65]{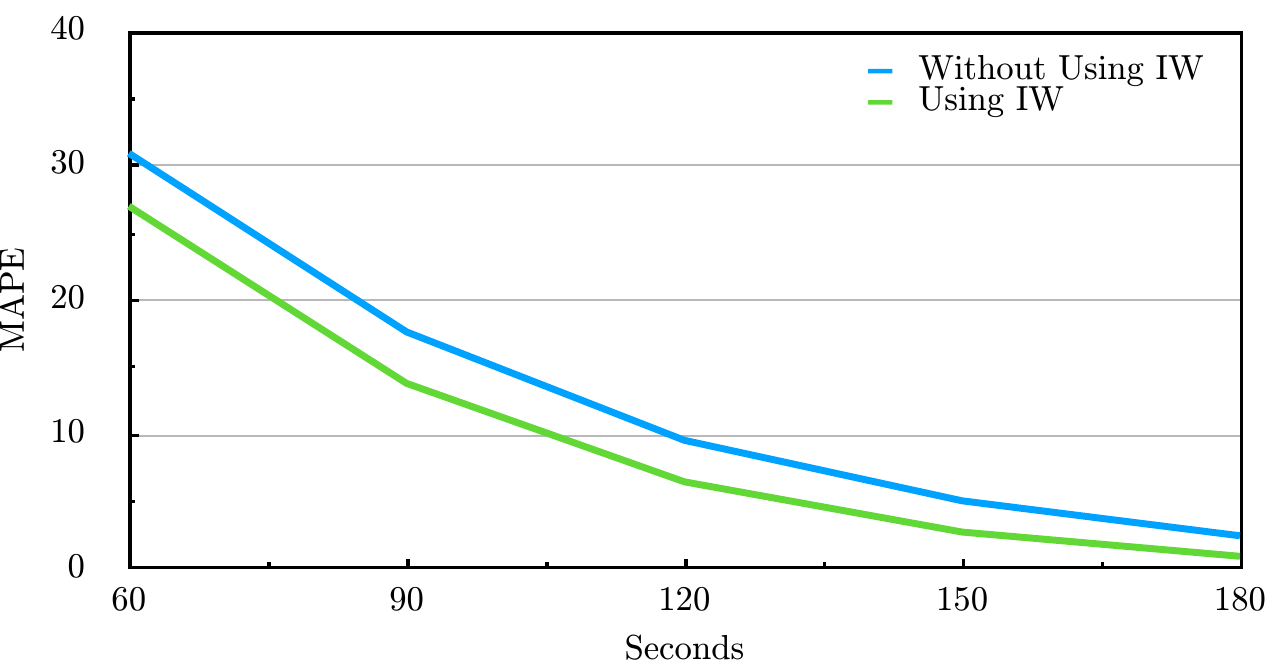}
    \caption{XGBoost results- Furnace A.}
    \label{fig:furnace_a_XG}
\end{subfigure}%
\hfill
\begin{subfigure}[b]{0.49\textwidth}
    \centering
    \includegraphics[scale=0.65]{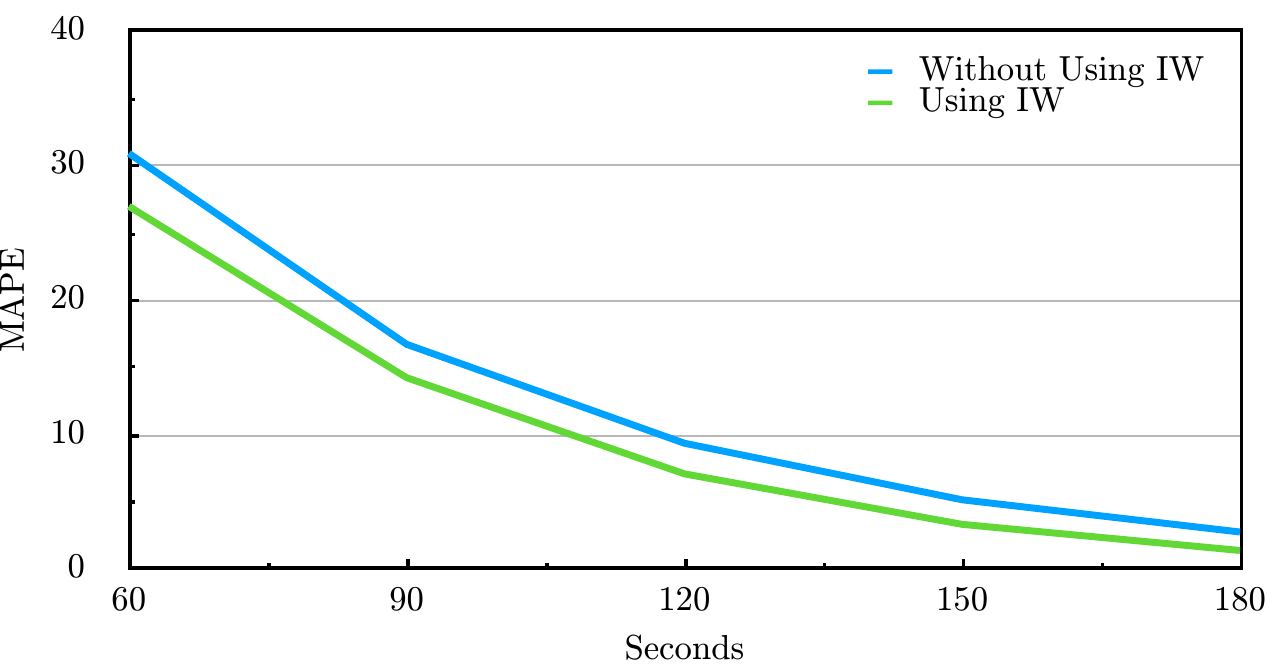}
    \caption{XGBoost results-Furnace B.}
    \label{fig:furnace_b_XG}
\end{subfigure}

\begin{subfigure}[b]{0.49\textwidth} 
    \centering
    \includegraphics[scale=0.65]{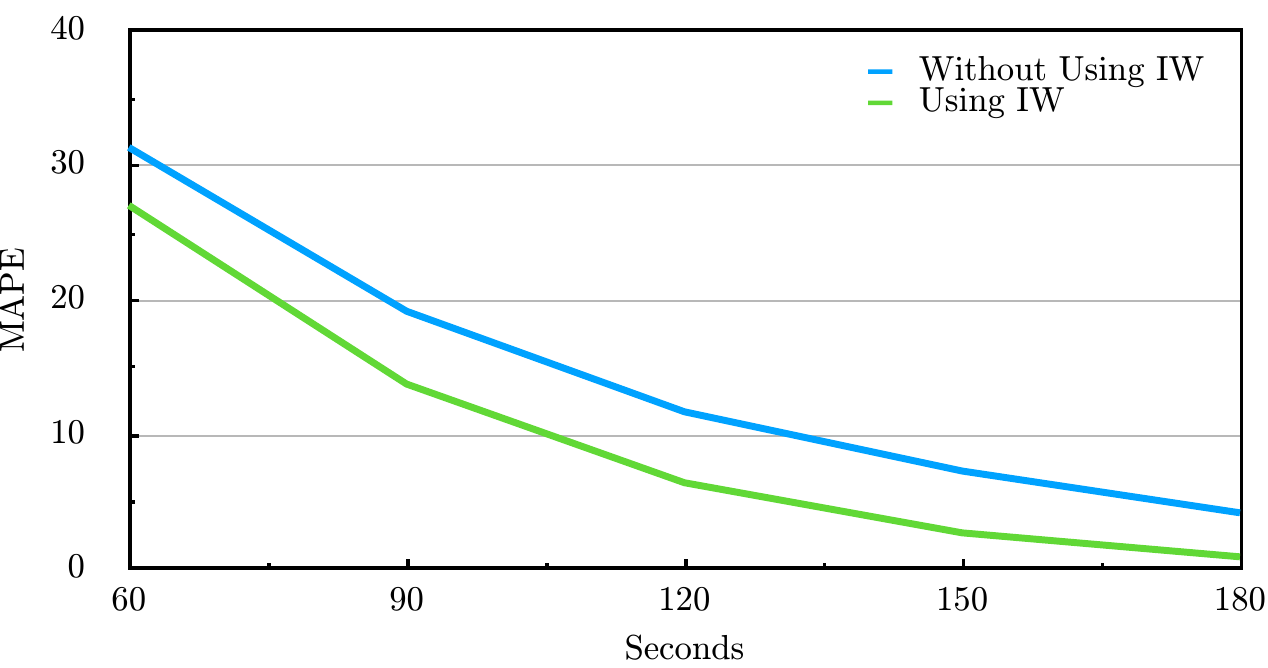}
    \caption{RF results- Furnace A.}
    \label{fig:furnace_a-RF}
\end{subfigure}%
\hfill
\begin{subfigure}[b]{0.49\textwidth} 
    \centering
    \includegraphics[scale=0.65]{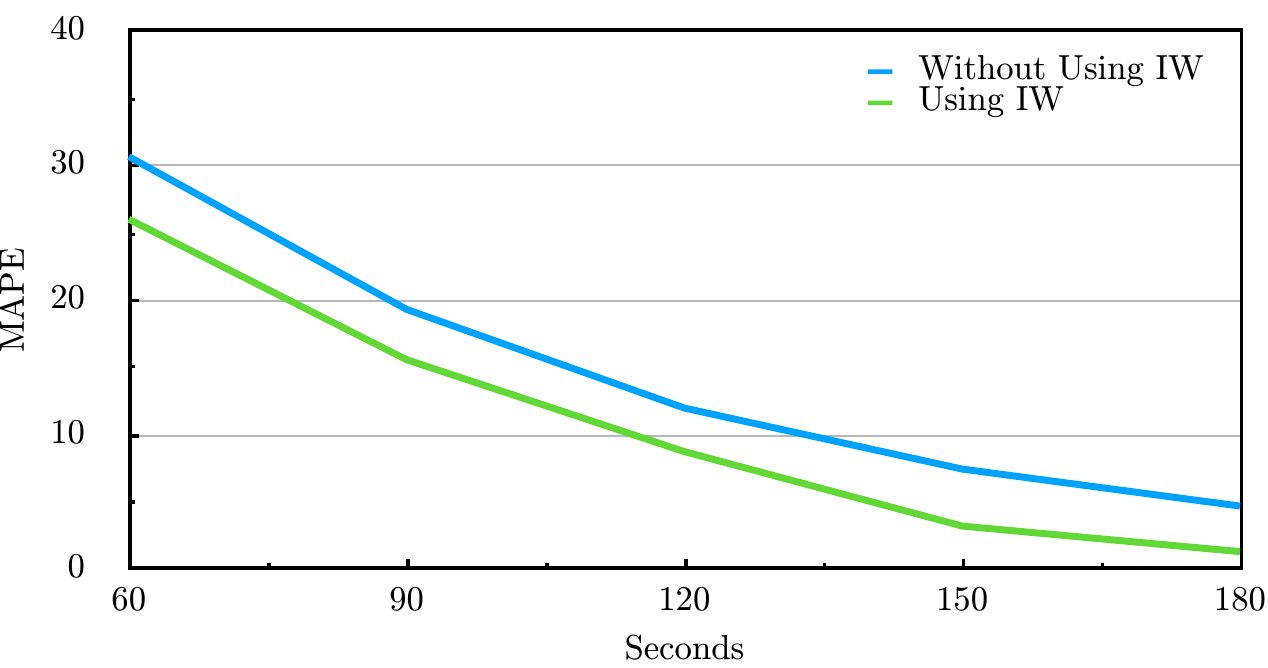}
    \caption{RF results-Furnace B.}
    \label{fig:furnace_b-RF}
\end{subfigure}%
    \caption{Evaluation results of the ESR use case} 
\label{fig:results}
\end{figure*}

Figures \ref{fig:furnace_a_XG} and \ref{fig:furnace_b_XG} compare the results of applying the XGBoost algorithm to the ESR data from the different furnaces. Figures \ref{fig:furnace_a-RF} and \ref{fig:furnace_b-RF} compare the results of applying the RF algorithm. The algorithms are trained on data collected from historical pumping events of furnace A and are evaluated on pumping events of both furnaces A and B. The results show the MAPE rates using different temporal intervals.  We can see an improvement in the error rate when the importance weighting is used in the four cases. More specifically, the error rate is lower when the models are evaluated in the furnace from where the pressure data were originally recorded. However, the error is slightly higher when we evaluate the model on new furnace datasets. This is because the data distribution of the new furnaces may not be as similar to the data distribution of the old furnace. Hence, fewer data points will have high-importance weight estimates following the KMM method. Usually, the error rates of the target domain are lower with the acquisition of more data points as the estimated PDF approaches the actual one.





We can also see a steady decline in the error rate as we enlarge the time window frame. As summarized in Tables \ref{tab:furnace_a_XG} and \ref{tab:furnace_a_RF}, for furnace A and without using the importance weighting technique, the MAPE rate starts at more than 30\% when using the pressure values of the first minute. Gradually, it decreases when the pressure values are used after the third minute to approximately 2.5\% when applying the XGBoost algorithm and approximately 4.2\% when applying the RF algorithm. However, when using the importance weighting technique, the MAPE rate starts approximately 27\%, reflecting an improvement of approximately 13\%, and decreases to less than 1\% after using the pressure values after the third minute. Approximately 61\% MAPE improvements for the XGBoost algorithm and approximately 78\% for the RF algorithm.

Analogously for furnace B and as revealed in Tables \ref{tab:furnace_b_XG} and \ref{tab:furnace_b_RF}, the MAPE rate starts at more than 30\% after the first minute without using the importance weighting technique. It reaches approximately 2.8\% for the XGBoost algorithm and 4.7\% for the RF algorithm. Using the importance weighting technique, the error starts approximately 27\% after the first minute and reaches around 1.4\% after the third minute for XGBoost, achieving approximately 13\% improvement. However, the error starts approximately 26\% with the RF algorithm after the first minute, reflecting an improvement of approximately 15\% and decreases to approximately 1.32\% after the third minute, reflecting an improvement of approximately 72\% of the MAPE evaluation metric.

Observations from the experimental evaluation showed that the application of an automated and adaptive solution yielded high improvement rates and solved the problem of lack of data when new conditions are introduced in industrial processes. The improvement is proportional to the data size in both ML algorithms used; i.e., the importance weighting methods perform better when fed more data points. Thus, the results demonstrate a trade-off between the predictive performance of the ML software and the decision-making time. Industry experts could determine the threshold that controls the trade-off by setting a minimum value of the performance metric to be achieved before signaling a maintenance alert.

\begin{table}
\centering
\small
\caption{Furnace A - XGBoost}
\label{tab:furnace_a_XG}
\begin{tabular}{llllll}
\hline
                        & \textbf{60} & \textbf{90} & \textbf{120} & \textbf{150} & \textbf{180}  \\ 
\hline
\textbf{Without IW} & 30.926      & 17.636      & 9.575        & 5.074        & 2.475         \\
\textbf{Using IW}         & 26.991      & 13.810      & 6.487        & 2.751        & 0.948         \\
\textbf{Improvement}      & 12.726\%   & 21.697\%   & 32.248\%    & 45.788\%    & 61.702\%     \\
\hline
\end{tabular}
\end{table}

\begin{table}
\centering
\small
\caption{Furnace A - RF}
\label{tab:furnace_a_RF}
\begin{tabular}{llllll}
\hline
                        & \textbf{60} & \textbf{90} & \textbf{120} & \textbf{150} & \textbf{180}  \\ 
\hline
\textbf{Without IW} & 31.369      & 19.190      & 11.718       & 7.314        & 4.209         \\
\textbf{Using IW}         & 27.061      & 13.778      & 6.445        & 2.720        & 0.938         \\
\textbf{Improvement}      & 13.732\%   & 28.204\%   & 44.998\%    & 62.818\%    & 77.723\%     \\
\hline
\end{tabular}
\end{table}

\begin{table}
\centering
\small
\caption{Furnace B - XGBoost}
\label{tab:furnace_b_XG}
\begin{tabular}{llllll}
\hline
                        & \textbf{60} & \textbf{90} & \textbf{120} & \textbf{150} & \textbf{180}  \\ 
\hline
\textbf{Without IW} & 30.926      & 16.733      & 9.381        & 5.176        & 2.783         \\
\textbf{With IW}         & 26.991      & 14.252      & 7.105        & 3.351        & 1.410         \\
\textbf{Improvement}      & 12.726\%   & 14.830\%   & 24.258\%    & 35.268\%    & 49.318\% \\
\hline
\end{tabular}
\end{table}

\begin{table}
\centering
\small
\caption{Furnace B - RF}
\label{tab:furnace_b_RF}
\begin{tabular}{llllll}
\hline
                        & \textbf{60} & \textbf{90} & \textbf{120} & \textbf{150} & \textbf{180}  \\ 
\hline
\textbf{Without IW} & 30.677      & 19.329      & 11.994       & 7.460        & 4.712         \\
\textbf{With IW}         & 26.033      & 15.589      & 8.757        & 3.224        & 1.324         \\
\textbf{Improvement}      & 15.138\%   & 19.350\%   & 26.985\%    & 56.787\%    & 71.909\%     \\
\hline
\end{tabular}
\end{table}

\section{Conclusion and Related Work}
Our paper shows an industrial use case of applying an automated approach that addresses scalable industrial processes. The approach is based on ML software that is able to satisfy the adaptability characteristic, which is considered a key non-functional requirement in evolving systems. We discussed the problem of upscaling the ESR process with a new furnace(s) in the Uddeholm steel company. The adaptive ML software uses a drift handling technique called importance weighting. The evaluation results show the improvement in the error rate of the adaptive solution over the non-adaptive solution. The adaptability of the proposed approach can open new opportunities for industries to integrate expandable solutions in scalable processes by reusing the knowledge extracted from previous similar tasks. 

In recent studies, importance weighting has been used in various domains to improve the performance of the predictive system. Rakotomamonjy \textit{et al.} \cite{rakotomamonjy2022optimal} has proposed a method that uses importance weighting based on the Wasserstein distance between the source and target domains and is applied to real-world visual domain adaptation problems. Similarly, in \cite{zhao2021active}, the authors have incorporated importance weighting techniques into deep active learning tasks and evaluated the proposed approach on image datasets. For the vacuum pumping problem, a recent study \cite{garg2021machine} has been presented to predict the minimum value of a vacuum pumping event. The exponentially weighted moving average (EWMA) chart has been integrated with the random forest algorithm in the predictive system. In a different approach \cite{chatterjee2022testing}, a data augmentation approach has been developed to generate more training samples using the principles of vacuum pumping to improve prediction performance.

In future work, it would be interesting to utilize the importance weights and incorporate them as one factor of the quality assurance mechanism, specifically the data quality component, of the overall MLOps system, in which the ML solution is eventually deployed. It would also be beneficial to test other importance weighting strategies and compare the results.   

\begin{acks}
This work has been funded by the Knowledge Foundation of Sweden (KKS) through the Synergy Project AIDA - A Holistic AI-driven Networking and Processing Framework for Industrial IoT (Rek:20200067).
\end{acks}

\bibliographystyle{ACM-Reference-Format}
\bibliography{sample-base}

\end{document}